\documentclass[12pt,a4paper]{article}
\usepackage{amsfonts,amsmath,amssymb,mathrsfs,amscd}
\title{A look inside the theory of the linear approximation}

\author{Ll.\ Bel\\
\emph{wtpbedil@lg.ehu.es}
}
\date{}
\begin{document}
\maketitle

\begin{abstract} 

We introduce in the framework of the linear approximation of General relativity a natural distinction between General gauge transformations generated by any vector field and those Special ones for which this vector field is a gradient. This allows to introduce geometrical objects that are not invariant under General gauge transformations but they are under Special ones. We develop then a formalism that strengthens the analogy of the formalisms of the electromagnetic and the gravitational theories in a Special relativity framework. We are thus able to define the energy-momentum tensor of the gravitational field and to fully analyze the gravitational field of isolated point masses or continuous distributions of them obtained by linear superpositions. 

\end{abstract}

\section*{Introduction}
This paper aims to a better understanding of the linear approximation of General relativity where the gravitational field is described by a space-time metric:

\begin{equation}
\label{0.1}
g_{\alpha\beta}=\eta_{\alpha\beta}+h_{\alpha\beta}
\end{equation}
where the symmetric deviations from Minkowski's metric are assumed to be small, and small varying, quantities. 

From the general theory that requires that the space-time metric has to be a tensor under general coordinate transformations it follows that $h_{\alpha\beta}$ are only defined up to a gauge transformation of the following type:

\begin{equation}
\label{0.2}
h_{\alpha\beta}\rightarrow h_{\alpha\beta}+\partial_\alpha\xi_\beta+\partial_\beta\xi_\alpha
\end{equation}  
$\xi_\alpha$ being any vector field. 

We define the rotational of $h_{\alpha\beta}$ as:

\begin{equation}
\label{0.3}
F_{\alpha\beta\mu}=\partial_\alpha h_{\beta\mu}-\partial_\beta h_{\alpha\mu}
\end{equation}
and the bi-rotational as: 

\begin{equation}
\label{0.4}
R_{\alpha\beta\mu\nu}=-\frac12(\partial_\mu F_{\alpha\beta\nu}-\partial_\nu F_{\alpha\beta\mu})
\end{equation}
or:

\begin{equation}
\label{0.4.1}
R_{\alpha\beta\mu\nu}=-\frac12(\partial_{\alpha\mu}h_{\beta\nu}+\partial_{\beta\nu}h_{\alpha\mu}-\partial_{\alpha\nu}h_{\beta\mu}-\partial_{\beta\mu}h_{\alpha\nu})
\end{equation}
which is of course the linear approximation of the Riemann tensor of the space-time \eqref{0.1}.
It is well-known that only this tensor, or tensors derived from it, depending on second order derivatives or higher of $h_{\alpha\beta}$, are invariant under General gauge transformations. On the contrary the rotational, that depends on first order derivatives, gets transformed according to:

\begin{equation}
\label{0.7}
F_{\alpha\beta\mu} \rightarrow F_{\alpha\beta\mu}+\partial_\mu(\partial_\alpha\xi_\beta-\partial_\beta\xi_\alpha)
\end{equation}
and it is invariant only in particular circumstances, as for example when the vector $\xi_\alpha$ is a gradient in which case we shall say that the gauge transformation is a Special gauge transformation.

The main idea of this paper consists in including as geometrical objects of physical interest those, like $F_{\alpha\beta\mu}$ and those derived from it, that are only invariant under Special gauge transformations. It will be organized as follows. In Sect.~1 we shall write down the field equations that $F_{\alpha\beta\mu}$ satisfies following from its definition and Einstein's field equations in vacuum. In Sect.~2 we shall define the energy-momentum tensor of the gravitational field. It is a 2-rank  tensor, quadratic in $F_{\alpha\beta\mu}$ that is  conserved in vacuum and that unlike the 4-rank super-energy tensor has the right physical dimensions. In Sect.~3 we apply in detail the preceding formalism to describe in full generality the gravitational field of an isolated point-mass moving along a geodesic world-line of Minkowski's space-time. In Sect.~4 we discuss the relevance of several gauge fixing conditions to simplify the geodesic equations. In Sect.~5 we consider a rigidly rotating constant-density disk and describe the asymptotic behavior of the stationary gravitational field obtained by a linear superposition of the fields of its infinitesimal constituents. In our Concluding remark we consider the possibility of going a little bit beyond Einstein's theory.

\section{Field equations}

We shall use the following terminology: we shall say that two potential tensors $h_{1\alpha\beta}$ and $h_{2\alpha\beta}$ are Weakly equivalent if they are equivalent modulo a General gauge transformation. A gravitational field can be defined thus as a class of Weakly equivalent potentials. We shall say that two potential tensors are Strongly equivalent if they are equivalent modulo a Special gauge transformation. We shall call a class of Strongly equivalent potentials a First order implementation of the corresponding gravitational field. Any gravitational field can be defined either by the bi-rotational $R_{\alpha\beta\lambda\mu}$ or of by the rotational $F_{\alpha\beta\lambda}$ of any particular First order implementation of it.

Let us consider any first order implementation $F_{\alpha\beta\lambda}$ of a gravitational field. From its very definition it follows that this field satisfies the constraint:

\begin{equation}
\label{3.0}
F_{[\alpha\beta\mu]}=0.
\end{equation}
where $[,]$ is the symbol for complete anti-symmetrization, and a first group of differential equations, namely:

\begin{equation}
\label{3.1}
\partial_{[\gamma} F_{\alpha\beta ]\mu}=0.
\end{equation}

On the other hand using (\ref{0.4}) we obtain:

\begin{equation}
\label{3.2}
\partial_\alpha F^\alpha_{\ \beta\mu}-\partial_\mu F_\beta=-2R_{\beta\mu}, \quad
\partial_\alpha F^\alpha=R,
\end{equation}
where $R_{\beta\mu}$ and $R$ are the linear parts of the Ricci tensor, and the curvature scalar:

\begin{equation}
\label{3.2.1}
R_{\beta\mu}=R_{\alpha\beta\lambda\mu}\eta^{\alpha\lambda}  \quad R=\eta^{\beta\nu}R_{\beta\nu},
\end{equation}
and where:

\begin{equation}
\label{3.3}
F_\beta=\eta^{\alpha\mu}F_{\alpha\beta\mu}.
\end{equation}
It follows then that:

\begin{equation}
\label{3.4}
\partial_\alpha H^\alpha_{\ \beta\mu}=-2S_{\beta\mu}
\end{equation}
where $S_{\beta\mu}$ is the linear part of the Einstein tensor:

\begin{equation}
\label{3.4.1}
S_{\beta\mu}=R_{\beta\mu}-\frac12 R\eta_{\beta\mu}
\end{equation}
and where:

\begin{equation}
\label{3.5}
 H_{\alpha\beta\mu}=F_{\alpha\beta\mu}+F_\alpha\eta_{\beta\mu}-F_\beta\eta_{\alpha\mu}
\end{equation}

Eqs. (\ref{3.1}) can also be written equivalently as: 

\begin{equation}
\label{3.8}
\partial_\alpha \stackrel{*}{F^\alpha}_{\ \beta\gamma}=0
\end{equation}
where $*$ is the dual operator acting on the pair of antisymmetric indices: 

\begin{equation}
\label{3.8.1}
\stackrel{*}F_{\alpha\beta\gamma}=
\frac12\delta_{\alpha\beta\mu\nu}^{0123}F^{\mu\nu}_{\ \ \gamma}
\end{equation}

Similarly Eq. (\ref{3.4}) can be written:

\begin{equation}
\label{3.9}
\partial_{[\gamma} \stackrel{*}{H}_{\alpha\beta]\mu}=2\eta_{\alpha\beta\gamma\rho}S^\rho_\mu                                                         
\end{equation}

\section{The Energy-momentum tensor}

Let us introduce the 2-rank tensor:

\begin{equation}
\label{3.14}
t_{\alpha\beta}=H_{\alpha\rho\mu} F^{\ \rho\mu}_{\beta}
+\stackrel{*}{F}_{\alpha\rho\mu}\stackrel{*}{H}^{\ \rho\mu}_{\beta\ \ }
\end{equation}
that can be written also as:

\begin{equation}
\label{3.14a}
t_{\alpha\beta}=2H_{\alpha\rho\mu}F^{\ \rho\mu }_{\beta\ \ }
-\frac12\eta_{\alpha\beta}{F}_{\sigma\rho\lambda}{H}^{\sigma\rho\lambda}
\end{equation}
This tensor is not symmetric in general and it is symmetric if and only if:

\begin{equation}
\label{3.14.3}
F^\rho F_{\alpha\beta\rho}=0.
\end{equation}

We start in this section the discussion of the properties that will make of a tensor proportional to it:

\begin{equation}
\label{3.14.0}
T_{\alpha\beta}=Bt_{\alpha\beta}
\end{equation}
$B$ being a pure numerical factor, a good candidate to be chosen as the energy-momentum density of the gravitational field in some particular First order implementations.

Using the field equations (\ref{3.1}) and (\ref{3.4}) we have:

\begin{equation}
\label{3.14.1}
\partial_\alpha(H^{\alpha\rho\mu}F_{\beta\rho\mu})=
\frac12 H^{\alpha\rho\mu}\partial_\beta F_{\alpha\rho\mu}-2S^{\rho\mu}F_{\beta\rho\mu}
\end{equation}
Similarly using (\ref{3.8}) and (\ref{3.9}) we have:

\begin{equation}
\label{3.14.2}
\partial_\alpha(\stackrel{*}{F}^{\alpha\rho\mu}\stackrel{*}{H}_{\beta\rho\mu})=
\frac12\stackrel{*}{F}^{\alpha\rho\mu}\partial_\beta\stackrel{*}{H}_{\alpha\rho\mu}
+4S^{\rho\mu}F_{\beta\rho\mu}
\end{equation}

Using now the identity:

\begin{equation}
\label{3.15}
\stackrel{*}{F}_{\alpha\rho\mu}\partial_\beta\stackrel{*}{H}^{\alpha\rho\mu}=
-F_{\alpha\rho\mu}\partial_\beta H^{\alpha\rho\mu}
\end{equation}
we obtain the following formula for the divergence of the tensor (\ref{3.14}):

\begin{equation}
\label{3.15.1}
\partial_\alpha t^\alpha_{\ \beta}=
\frac12(H^{\alpha\rho\mu}\partial_\beta F_{\alpha\rho\mu}-F^{\alpha\rho\mu}\partial_\beta H_{\alpha\rho\mu})
+2S^{\rho\mu}F_{\beta\rho\mu}
\end{equation}
and remembering the definition (\ref{3.5}) we get:

\begin{equation}
\label{3.16}
\partial_\alpha t^\alpha_{\ \beta}=-\partial_\beta(F^\alpha F_\alpha)+2F^\alpha\partial_\beta F_\alpha+2S^{\rho\mu}F_{\beta\rho\mu}
\end{equation}
or, simplifying: 

\begin{equation}
\label{3.18}
\partial_\alpha t^\alpha_{\ \beta}=2S^{\rho\mu}F_{\beta\rho\mu}
\end{equation}
and therefore $t_{\alpha\beta}$ is conserved when Einstein's field equations in vacuum:

\begin{equation}
\label{3.18.1}
S_{\alpha\beta}=0
\end{equation}
are satisfied.

Contrary to this latter property that it is independent of the First order implementation being used the value of the time-component:

\begin{equation}
\label{3.19}
W(e)=t_{\alpha\beta}e^\alpha e^\beta 
\end{equation}
where $e^\alpha$ is any time-like vector, will depend on the implementation and has to be discussed in each case. We shall do it in detail in one important case in the next section.
 
Let us consider, as a simple example, a gravitational plane-wave propagating in the $x^3$ direction. The familiar harmonic and trace free corresponding potential tensor is:

\begin{equation}
\label{3.20}
h_{\alpha\beta}=A(u)(\delta_{1\alpha}\delta_{1\beta}-\delta_{2\alpha}\delta_{2\beta}), \quad
u=t-x^3
\end{equation}
and the corresponding expression of the First order implementation of the gravitational field is:

\begin{equation}
\label{3.21}
F_{\gamma\alpha\beta}=\left(\frac{d}{du}\ln A\right)\left((\delta_{0\gamma}-\delta_{3\gamma})h_{\alpha\beta}-
(\delta_{0\alpha}-\delta_{3\alpha})h_{\gamma\beta}\right)
\end{equation}    
In this case we have:

\begin{equation}
\label{3.22}
F_\alpha=0, \quad H_{\gamma\alpha\beta}=F_{\gamma\alpha\beta}, \quad 
F_{\rho\alpha\beta}F^{\rho\alpha\beta}=0
\end{equation}
and therefore the tensor (\ref{3.14}) is simply:

\begin{equation}
\label{3.23}
t_{\alpha\beta}=F_{\alpha\rho\sigma}F^{\ \rho\sigma}_\beta
\end{equation}
It leads to the following result:

\begin{equation}
\label{3.24}
t_{\alpha\beta}=2\left(\frac{dA}{du}\right)^2l_\alpha l_\beta, \quad l_\rho=\partial_\rho u
\end{equation}
and encourages the interpretation of (\ref{3.14}) as being proportional to the energy-momentum tensor of the plane wave.

\section{The gravitational field of a point mass}

As early as in 1906 Poincar\'e \cite{Poincare} used the Special relativity framework to propose a theory of the gravitational field of a point mass. In 1922 Whitehead \cite{Whitehead}\,\footnote{Reference \cite{Coleman} contains also abundant information about Whitehead's theory}, a few years after Einstein's General relativity was successfully completed and tested, reconsidered a point of view similar to that of Poincar\'e to develop what he thought could be a concurrent theory of General relativity avoiding some aspects of this theory that he disliked.

Let us consider a point particle with mass $m$ moving with constant 4-velocity $u^\beta$ along a time-like world-line $\cal W$ of Minkowski's space-time, and let $x^\alpha$ be the coordinates of a generic event where one wants the gravitational field created by $m$ to be calculated. We shall use the following notations: $\hat x^\alpha$ will be the coordinates of the intersection of the past light-cone $\cal C_-$ issued from $x^\alpha$ with $\cal W$; $L^\alpha$ will be the null vector:

\begin{equation}
\label{1.1}
L^\alpha=x^\alpha-\hat x^\alpha;
\end{equation}
and $r\geq 0$ will be the scalar:

\begin{equation}
\label{1.2}
r=-u_\alpha L^\alpha
\end{equation}
which is the space distance from $x^\alpha$ to $\cal W$. Actually we shall use the reduced null vector $l^\alpha$:

\begin{equation}
\label{1.3}
l^\alpha=\frac{1}{r}L^\alpha
\end{equation}
instead of $L^\alpha$.
The algebraic equalities:

\begin{equation}
\label{1.4}
u_\alpha u^\alpha=-1, \quad u_\alpha l^\alpha=-1, \quad l_\alpha l^\alpha=0,
\end{equation}
together with the easily derived formulas below are the main side conditions to be taken into account in the calculations to be made in this section: 

\begin{equation}
\label{1.5}
\partial_\rho u_\alpha=0, \quad \partial_\rho r=n_\rho, \quad 
\partial_\rho l_\alpha=\frac{1}{r}(\eta_{\alpha\rho}+u_\rho l_\alpha+l_\rho u_\alpha-
l_\rho l_\alpha)
\end{equation}
where $\partial_\rho$ means partial derivative with respect to $x^\rho$ and where:

\begin{equation}
\label{1.5.1}
n_\rho=-u_\rho+l_\rho
\end{equation}

Whitehead's theory is based on the assumption that the gravitational field of the point-particle $m$ is described by the following symmetric covariant tensor invariant by the Poincar\'e group:

\begin{equation}  
\label{1.6}
g_{\alpha\beta}=\eta_{\alpha\beta}-\frac{2m}{r}l_\alpha l_\beta
\end{equation}
where we have assumed that $G=c=1$. 

From Whitehead's point of view (\ref{1.6}) is the foundation of a new theory but in fact (\ref{1.6}) is the exact Schwarzschild solution of Einstein's equations, written in a particular system of coordinates that nowadays are called Kerr-Schild coordinates.

This section generalizes the starting point of Whitehead while sticking to the point of view that we are dealing with the linear approximation of Einstein's theory. We assume thus 
Eq.~(\ref{0.1}) requiring
$h_{\alpha\beta}$ to be the more general Poincar\'e invariant tensor built with the geometrical quantities at our disposal, i.e, $l^\alpha$, $u^\alpha$ and four constants $A_0$-$A_3$ with the dimension of length in the system of units that we are using:

\begin{equation}
\label{2.2}
h_{\alpha\beta}=\frac{1}{r}\left(A_0u_\alpha u_\beta+A_1\eta_{\alpha\beta}
+A_2(u_\alpha l_\beta+l_\alpha u_\beta)-A_3l_\alpha l_\beta\right)
\end{equation}
 
To maintain $h_{\alpha\beta}$ as a Poincar\'e invariant tensor the more general acceptable gauge transformation that we shall consider is that where:

\begin{equation}
\label{2.9}
\xi_\alpha=\frac12 k_1 l_\alpha + k_2 u_\alpha\ln r
\end{equation}
$k_1$ and $k_2$ being two constants.
From the third set of Eqs.~(\ref{1.5}) it follows that $l_\alpha$ is a gradient and therefore if $k_2=0$ the transformation is a Special gauge transformation as defined in the Introduction.

A General gauge transformation such as (\ref{2.9})induces the following transformation on the coefficients $A_0$-$A_3$ in (\ref{2.2}):

\begin{equation}
\label{2.9.1}
A_0\rightarrow A_0-2k_2, \quad A_1\rightarrow A_1+k_1, \quad
A_2\rightarrow A_2+k_1+k_2, \quad A_3\rightarrow A_3+k_1
\end{equation}

Let us consider a linear combination of the coefficients $A_0$-$A_3$:

\begin{equation}
\label{2.10}
C=\alpha_0A_0+\alpha_1A_1+\alpha_2A_2+\alpha_3A_3
\end{equation}
Necessary and sufficient conditions that leave $C$ invariant are the following:

\begin{equation}
\label{2.11}
\alpha_1+\alpha_2+\alpha_3=0,\quad -2\alpha_0+\alpha_2=0
\end{equation}
Some invariant linear combinations have very special tensor meanings. For example we have:

\begin{equation}
\label{2.12}
A_0-2(A_1-A_2)=0 \Leftrightarrow R_{\alpha\mu}=0
\end{equation}
and:

\begin{equation}
\label{2.13}
\{A_0-2(A_1-A_2)=0, \quad \hbox{and} \quad A_1-A_3=0\} \Leftrightarrow R_{\alpha\beta\mu\nu}=0
\end{equation}

Let us consider a frame of reference such that: 

\begin{equation}
\label{2.14}
u^0=1, \quad u^i=0, \quad \hat x^i=0, 
\end{equation}
Then we have:

\begin{equation}
\label{2.15}
l^0=1, \quad l^i=\frac{x^i}{r}, \quad r^2=\sum (x^k)^2
\end{equation}

Using (\ref{2.2}) the components of the metric (\ref{0.1}) become:

\begin{eqnarray}
\label{2.16a}
g_{00}&=&-1+\frac{A_0-A_1+2A_2-A_3}{r} \\
\label{2.16b}
g_{0i}&=&\qquad  - \frac{A_2-A_3}{r^2}x_i\\
\label{2.16c}
g_{ij}&=&\ \delta_{ij}+\frac{A_1r^2\delta_{ij}-A_3 x_ix_j}{r^3}
\end{eqnarray}

An obvious geometrical invariant in the corresponding stationary frame of reference at the linear approximation is:
\begin{equation}
\label{2.16.1}
\sqrt{-g_{00}}=1-\frac{A_0-A_1+2A_2-A_3}{2r}
\end{equation}
that becomes using (\ref{2.12}) and assuming the vacuum field equations (\ref{3.18.1}), as we shall do from now on unless otherwise stated:

\begin{equation}
\label{2.16.1.1}
\sqrt{-g_{00}}=1-\frac{A_1-A_3}{2r}
\end{equation}
From it we derive the force per unit test mass at rest with respect to the point source:

\begin{equation}
\label{2.16.2}
f_i=-\partial_i\sqrt{-g_{00}}=-\frac{A_1-A_3}{2r^2}
\end{equation}
leading to the forced identification:

\begin{equation}
\label{2.16.2.1}
A_1-A_3=2m
\end{equation}
to guarantee the appropriate Newtonian limit. 

If $A_1=A_2=0$ (\ref{2.16a})-(\ref{2.16c}) is the Schwarzschild solution, in Kerr-Schild coordinates, as we mentioned before. If $A_2=A_3$ and $A_1=0$ we have the linearized Schwarzschild solution written in curvature coordinates, and if $A_2=A_3=0$ and $A_1=2m$ we have it in harmonic, or equivalently at this approximation, isotropic coordinates.

More generally it can be seen easily that any two vacuum solutions with the same value of $m$ describe the same gravitational field because their potentials are necessarily Weakly equivalent. This is a simplified version of Birkoff's theorem at the linear approximation. 

The remaining invariant geometrical objects are the Rotation rate field, which is zero:

\begin{equation}
\label{2.16.2.2}
\Omega_{ij}=\partial_i h_{0j}-\partial_j h_{0i}=0
\end{equation}
and the linear approximation of the curvature tensor of the space metric $g_{ij}$ above which is:

\begin{equation}
\label{2.16.3}
R_{ikjl}=\frac{m}{r^3}(2(\delta_{ij}\delta_{kl}-\delta_{kj}\delta_{il})+
3(\delta_{ij}l_kl_l+\delta_{kl}l_il_j-\delta_{il}l_kl_j-\delta_{kj}l_il_l))
\end{equation}

Using spherical polar coordinates we may write (\ref{2.16a})-(\ref{2.16c}) as the manifestly spherically symmetric line-element:

\begin{equation}
\label{2.18}
ds^2=-\left(1-\frac{2m}{r}\right)dt^2-2\frac{A_2-A_3}{r}drdt+\left(1+\frac{2m}{r}\right)dr^2+\left(1+\frac{A_1}{r}\right)r^2d\Omega^2
\end{equation}

Keeping in mind (\ref{2.12}) and using (\ref{2.2}) a straightforward calculation yields:

\begin{eqnarray}
\nonumber
F_{\alpha\beta\lambda}=\frac{1}{r^2}(2(A_2-A_3)l_\lambda+2(A_1-A_2)u_\lambda)(u_\alpha l_\beta-u_\beta l_\alpha) \\  
\label{2.18.1}
-\frac{1}{r^2}(\eta_{\lambda\alpha}((A_1-A_2)u_\beta-2ml_\beta)-\eta_{\lambda\beta}((A_1-A_2)u_\alpha-2ml_\alpha))
\end{eqnarray}
from which expression and (\ref{3.14a}) we obtain by a straightforward calculation:

\begin{equation}
\label{2.19}
t_{\alpha\beta}=\frac{2K}{r^4}(2l_\alpha l_\beta-2(l_\alpha u_\beta+l_\beta u_\alpha)-\eta_{\alpha\beta})
\end{equation}
where:

\begin{equation}
\label{2.19.1}
K=(A_1-A_2)(2m+A_2-A_3)
\end{equation}
As expected (\ref{2.19}) is not invariant under General gauge transformations. Nevertheless the gauge dependence does not come from the tensor structure, but from the scalar $K$, which actually can take any value including zero. 

Let us consider a unit time-like vector $e^\alpha$ that we can always write as:

\begin{equation}
\label{2.19.2}
e^\alpha=a_uu^\alpha+a_ll^\alpha+a_pp^\alpha
\end{equation}
where $p^\alpha$ is a unit space-like vector orthogonal to $u^\alpha$ and $l^\alpha$. Then we have, using (\ref{3.14.0}):

\begin{equation}
\label{2.19.3}
T_{\alpha\beta}e^\alpha e^\beta=\frac{2BK}{r^4}(1+2a_p^2)
\end{equation}
And in particular the time component of the tensor is, when the point mass is at rest:

\begin{equation}
\label{2.20}
W(u)=T_{\alpha\beta}u^\alpha u^\beta=\frac{2BK}{r^4}
\end{equation}
while the gravitational energy density at the Newtonian approximation is:

\begin{equation}
\label{2.21}
W_N=-\frac{1}{8\pi}\frac{m^2}{r^4}
\end{equation}
It is always possible to choose a First order implementation such that $W(u)=W_N$. In fact with a Gauge transformation (\ref{2.9}) with $k_2=A_3-A_2$, whatever be the initial values, the new values of the $A$`s are:

\begin{equation}
\label{2.22}
A_1\rightarrow A_1, \quad A_2\rightarrow A_3 \quad A_3\rightarrow A_3
\end{equation}
and therefore using (\ref{2.16.2.1}):

\begin{equation}
\label{2.23}
K\rightarrow (A_1-A_3)^2=4m^2
\end{equation}
Choosing then the numerical factor:

\begin{equation}
\label{2.23.1}
B=-\frac{1}{64\pi}
\end{equation}
we obtain the tensor: 

\begin{equation}
\label{2.24}
T_{\alpha\beta}=\frac{m^2}{8\pi r^4}(2l_\alpha l_\beta-2(l_\alpha u_\beta+l_\beta u_\alpha)-\eta_{\alpha\beta})
\end{equation}
that in our opinion is well qualified to be the definition of the energy-momentum tensor density of the gravitational field of a point mass and that, taken as a definition, is independent of any first order implementation.

This tensor could have been obtained directly from the following gauge invariant properties: i) It is a symmetric Poincar\'e invariant tensor built with $l_\alpha$ and $u_\alpha$; ii) it has the right physical dimensions; iii) it is conserved; iv) the corresponding energy density is definite negative and has the correct Newtonian limit; v) last but not least: there is a class of first order implementations such that it coincides with (\ref{3.14}) up to the numerical factor (\ref{2.23.1})

\section{The geodesic equations}

The geodesic equations of a test massive particle or a massless photon, in the gravitational field of a point mass described by the space-time metric (\ref{2.18}) are:

\begin{equation}
\label{2.25}
\frac{d^2x^\alpha}{d\lambda^2}
+\Gamma^\alpha_{\mu\nu}\frac{dx^\mu}{d\lambda}\frac{dx^\nu}{d\lambda}
=b\frac{dx^\alpha}{d\lambda}
\end{equation}
where the function $b$ depends on the parameter $\lambda$ chosen, and where the non zero Christoffel symbols are:

\begin{eqnarray}
\label{2.26a}
\Gamma^0_{01}&=&\frac{m}{r^2} \\
\label{2.26b}
\Gamma^1_{00}&=&\frac{m}{r^2}, \quad \Gamma^1_{11}=-\frac{m}{r^2}, \quad \Gamma^1_{22}=-r-\frac{A_1-2m}{2}, \quad \Gamma^1_{33}=\Gamma^1_{22}\sin^2\theta \\
\label{2.26c}
\Gamma^2_{12}&=&\frac1r-\frac{A_1}{2r^2}, \quad  \Gamma^2_{33}=-\sin\theta\cos\theta \\
\label{2.26d}
\Gamma^3_{13}&=&\frac1r-\frac{A_1}{2r^2}, \quad \Gamma^3_{23}=\frac{\cos\theta}{\sin\theta}
\end{eqnarray}

As we see, some of these symbols depend on a free parameter $A_1$. This is how the covariance of General relativity is reflected at the linear approximation. This means that full operational predictions will have to use these equations more than once to make it intrinsic. Most often this will require to use the geodesic equations above to describe the motion of test massive particles in conjunction with the same equations to describe the propagation of light. In which case, in the description of the experiment, $A_1$ can be chosen at our convenience depending on each particular case. In other circumstances, when a full operational description is impossible or too difficult to implement the theory may recur to some supplementary assumption. The simplest one when the velocity of the particles is very small is that of the Newtonian approximation that assumes that the gravitational field does not affect the propagation of light and that the coordinates $x^i$ mean what they mean in Euclidean geometry. But less drastic assumptions may be used.

We see from (\ref{2.16a})-(\ref{2.16c}) that if the problem suggests using Cartesian coordinates of space then $A_1=2m$ is very likely the most convenient choice. On the other hand if the problem suggests using spherical polar coordinates then $A_1=0$ is the choice that makes the Christoffel symbols simpler.

\section{Extended bodies: an example}

An important benefit of dealing with a linear theory and having a complete model of the gravitational field of a point mass is that by linear superposition of the gravitational field of each of its components we can gain access to a variety of more complex sources, without needing matching conditions nor prescribing the appropriate asymptotic behavior. Synge \cite{Coleman} considered the case of a homogeneous spherically symmetric source and here we shall sketch another example that includes rotation and non sphericity of the source. 

We consider a flat disk of radius $a$ with constant surface mass density $\sigma$ and mass $M$, rigidly rotating with angular velocity $\omega$ on its fixed plane and center $O$. $y^i$ will be the Cartesian coordinates of the location $Y$ of a generic element of the disk, $y^3=0$ corresponding to the plane of the disk and $y^i=0$ to its center; $x^i$ will be the coordinates of the location $X$ exterior to the disk where the gravitational field is calculated at time $t$. We shall also use polar spherical coordinates $R,\theta,\phi$  and plane polar coordinates $\rho,\alpha$ on the plane of the disk. The square of the space distance from $Y$ to $X$ will be:

\begin{equation}
\label{4.1}
s^2=R^2\left((\sin\theta\cos\phi-(\rho/R)\cos\alpha)^2
+(\sin\theta\sin\phi-(\rho/R)\sin\alpha)^2+\cos\theta^2\right)
\end{equation}

The velocity components of any infinitesimal component of the disk located at $Y$ will be:

\begin{equation}
\label{4.2}
v^1=-\omega y^2= -\omega\rho\sin\alpha, \quad v^2=+\omega y^1= \omega\rho\cos\alpha
\end{equation}

We shall assume now that these quantities are small of order $\epsilon$, say, as well as the quantity $(\rho/R)^2$. Under such assumptions, neglecting the terms of order smaller than $\epsilon$, the components of $u_\alpha$ and $l_\beta$ at this order are:

\begin{equation}
\label{4.3}
u_0=-1, \quad u_i=v_i, \quad l^0=1+v_jn^j, \quad l_i=l^0n_i
\end{equation}
where $n_i$ is the unit vector pointing from $Y$ to $X$. At the same order we have also:

\begin{equation}
\label{4.4}
\frac1r=\frac1R\left(1+\left(\frac{\rho}{R}\cos(\alpha-\phi)-\omega\rho\sin(\alpha-\phi)\right)\sin\theta\right) 
\end{equation} 

With these assumptions and using the gauge fixing conditions $A_2=A_3=0$ we have, following the usual procedure to transform a discrete sum into a continuous distribution, the following leading terms:      

\begin{eqnarray}
\label{4.5a}
h_{00}&=&\ \frac{2M}{R}\left(1+\frac{a^2}{4R^2}(1-3\cos^2\theta)\right)\\
\label{4.5b}
h_{01}&=&\ \frac{M\omega a^2}{R^3}x^2 \\
\label{4.5c}
h_{02}&=&-\frac{M\omega a^2}{R^3}x^1 \\
\label{4.5d}
h_{03}&=&\ 0 \\
\label{4.5e}
h_{ij}&=&\ \frac{2M}{R}\left(1+\frac{a^2}{4R^2}(1-3\cos^2\theta)\right)\delta_{ij}
\end{eqnarray}
that correspond to an approximate solution with a monopole-quadrupole rotating source.

\section{Concluding Remark}

Maxwell equations in vacuum for a point charge $e$ moving with arbitrary acceleration:

\begin{equation}
\label{5.1}
\xi^\alpha= \frac{du^\alpha}{d\tau}
\end{equation}
where $\tau$ is proper Minkowski's time, follow from the assumption that the electromagnetic field is the rotational of a potential of the following form:

\begin{equation}
\label{5.2}
\phi_\alpha= \frac{1}{r}(B_1u_\alpha-B_3(1+r\hat\xi^\rho l_\rho)l_\alpha) 
\end{equation}
with $r$ and $l_\alpha$ as defined in (ref{1.2}) and (ref{1.3}), $\hat\xi^\rho$ being the acceleration at the retarded point $\hat x^\alpha$, and:

\begin{equation}
\label{5.3}
B_1-B_3=e
\end{equation} 
We may wonder then why we needed Einstein's equations in vacuum to describe the gravitational field of a point mass. The answer is that we do not need them, at least at the linear approximation, if we are ready to deal with a somewhat more general theory. In fact, if we abandon the condition (\ref{2.12}), from (\ref{2.16.1}) it follows that to maintain the Newtonian limit, we have to make the identification:

\begin{equation}
\label{5.4}
A_0-A_1+2A_2-A_3=2m
\end{equation}
and then writing:

\begin{equation}
\label{5.5}
A_1-A_3=2\gamma m
\end{equation}
gamma can be considered as a second fundamental constant on which depends the gravitational field. This $\gamma$ can be identified with Edington's parameter of the PPN formalism \cite{Will} when using the gauge fixing condition $A_3=0$

Keep in mind though that this would lead to a difficult interpretation of the Einstein's tensor:

\begin{equation}
\label{5.6}
S_{\alpha\beta}=(1-\gamma)\frac{2m}{r^3}(\eta_{\alpha\beta}+u_\alpha u_\beta-3n_\alpha n_\beta)
\end{equation} 
where we have used the notation (\ref{1.5.1}). 

\section*{Acknowledgments} 

I gratefully acknowledge a careful reading of the preceding version of this manuscript by J. Mart\'{i}n and A. Molina.

\end{document}